\documentclass[BCOR0mm,DIV13,12pt,a4paper,english]{scrartcl}
\pdfoutput=1
\usepackage{hgbreport}
\reportfooter{\scriptsize{\textsf{%
\hspace{-10mm}Applying recent secure element relay attack scenarios to the real world:\\%
\hspace{-10mm}Google Wallet Relay Attack\\[4pt]%
\hspace{-10mm}Michael Roland $<$\href{mailto:michael.roland@fh-hagenberg.at}{michael.roland@fh-hagenberg.at}$>$%
\hfill\textbf{}\\%
}}}

\RequirePackage[sort]{natbib}\bibpunct{ [}{]}{,}{n}{}{,}

\hypersetup{
  pdftitle={Technical Report: Applying recent secure element relay attack scenarios to the real world: Google Wallet Relay Attack},
  pdfauthor={Michael Roland; NFC Research Lab Hagenberg},
  pdfsubject={},
  pdfkeywords={}
}

\newcommand{\nohyph}{\hspace{0em}}

\begin{document}
\selectlanguage{english}

\title{%
  {\Large Technical Report}\\[-6pt]%
	\vspace{0.5cm}%
	{Applying recent secure element relay attack scenarios to the real world:\\Google Wallet Relay Attack}}
\author{%
  {\large Michael Roland}\\[-0pt]%
  {\normalsize NFC Research Lab Hagenberg}\\[-8pt]%
  {\normalsize University of Applied Sciences Upper Austria}\\[-8pt]%
  {\normalsize \href{mailto:michael.roland@fh-hagenberg.at}{michael.roland@fh-hagenberg.at}}}
\date{}
\maketitle

\begin{center}
	\begin{minipage}{0.7\textwidth}
	\subsubsection*{Abstract}
		This report explains recent developments in relay attacks on contactless smartcards
		and secure elements. It further reveals how these relay attacks can be applied to
		the Google Wallet. Finally, it gives an overview of the components and results of a
		successful attempt to relay an EMV Mag-Stripe transaction between a Google Wallet
		device and an external card emulator over a wireless network.
%	\subsubsection*{Keywords}
%		Relay attack, Google Wallet, Secure Element, Near Field Communication
	\end{minipage}
\end{center}

\begin{table}[!b]
\begin{tabular*}{\textwidth}{p{30mm}l}
\toprule[2pt]
\emph{Revision:}        & 1.3\\\midrule[0.75pt]
\emph{Date:}            & March 25, 2013\\\midrule[0.75pt]%August 27, 2012%\today\\\midrule[0.75pt]
\emph{Status:}          & Final\\\midrule[2pt]%\midrule[0.75pt]
\multicolumn{2}{p{.97\textwidth}}{\footnotesize This work is part of the project ``4EMOBILITY'' within the EU programme ``Regionale Wett\-be\-werbs\-fäh\-ig\-keit OÖ 2007--2013 (Regio 13)'' funded by the European regional development fund (ERDF) and the Province of Upper Austria (Land Oberösterreich).}\\%\bottomrule[2pt]%\midrule[0.75pt]
%\emph{Classification:}  & Unclassified\\\bottomrule[2pt]
\end{tabular*}
\end{table}

\begin{table}[!p]
\centering
\setlength{\fboxsep}{10pt}\setlength{\fboxrule}{3pt}
\small
\fbox{
\begin{minipage}{0.80\textwidth}
\begin{center}
A revised version of this report has been published as
\end{center}

\textsc{M. Roland, J. Langer, and J. Scharinger}: \emph{Applying Relay Attacks to Google Wallet}. In: Proceedings of the 5th International Workshop on Near Field Communication (NFC 2013), pp.\ 1--6, Zurich, Switzerland, Feb.\ 2013. DOI: \href{http://dx.doi.org/10.1109/NFC.2013.6482441}{10.1109/NFC.2013.6482441}, \copyright~2013 IEEE.

\vspace{15pt}The revised version gives a more detailed analysis of the Google Wallet on-card component, adds new ideas on how to improve the attack, details further information on our test setup, and adds an analysis of Google's approach to fix the vulnerability.
\end{minipage}
}
\end{table}

\clearpage\phantomsection\label{front:toc}\pdfbookmark[1]{\contentsname}{front:toc}
\tableofcontents
%\listoftables
%\listoffigures
\clearpage

\section{Introduction}
Recently, there have been several publications on relay attacks on contactless smartcards and secure element-enabled mobile devices\cite{Hancke2005,Kfir2005,Hancke2009,Francis2010,Francis2011,Roland2012,Roland2012IFIP}. However, relay attacks on contactless smartcards are not new. In 2005, Hancke\cite{Hancke2005} first showed that it is possible to relay contactless smartcard communication (ISO/IEC 14443\cite{iso14443}) over longer distances through an alternative communication channel. Kfir and Wool\cite{Kfir2005} describe a similar system and also show that the relay device used to access a victims card can be up to 50 centimeters away from the card when using additional amplification and filtering.

\subsection{Relay Attack}
\begin{figure}[!ht]
  \centering
  \subfigure[without relay]{%
            \includegraphics[height=32mm]{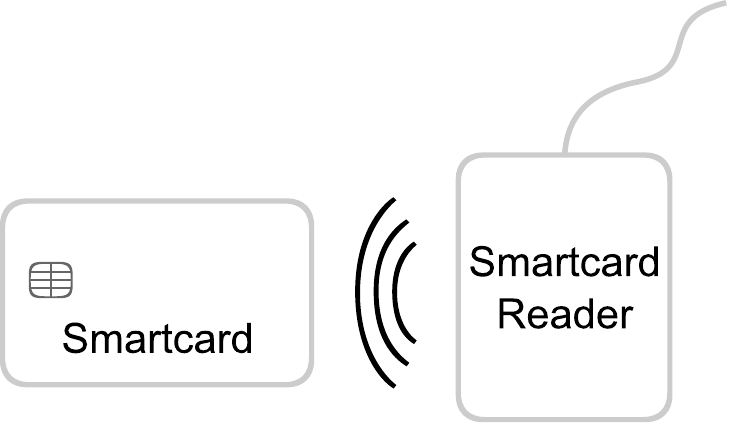}%
            \label{fig:relay+schematic+norelay}}\\
  \subfigure[with relay]{%
            \includegraphics[height=32mm]{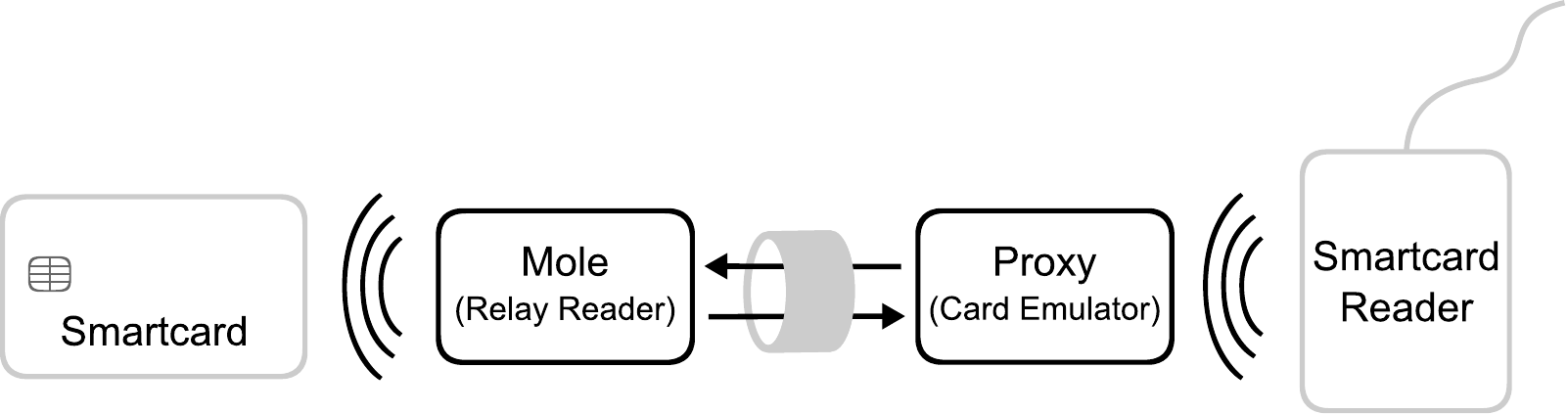}%
            \label{fig:relay+schematic+relay}}
  \caption{Communication between a smartcard and a reader.}%
  \label{fig:relay+schematic}
\end{figure}

A relay attack can be seen as a simple range extension of the contactless communication channel (see Fig.~\ref{fig:relay+schematic}). Thus, an attack requires three components:
\begin{enumerate}
	\item a reader device (often called \emph{mole}\cite{Hancke2005} or \emph{leech}\cite{Kfir2005}) in close proximity to the card under attack,
	\item a card emulator device (often called \emph{proxy}\cite{Hancke2005} or \emph{ghost}\cite{Kfir2005}) that is used to communicate with the actual reader, and
	\item a fast communication channel between these two devices.
\end{enumerate}
The attack is performed by bringing the mole in proximity to the card under attack. At the same time, the card emulator is brought into proximity of a reader device (POS terminal, access control reader...) Every command that the card emulator receives from the actual reader is forwarded to the mole. The mole, in turn, forwards the command to the card under attack. The card's response is then received by the mole and sent all the way back through the card emulator to the actual reader.

This type of attack cannot be prevented by application-level cryptography (e.g.\ encryption)\cite{Hancke2005,Hancke2009}. The problem is, that the relay attack is a simple range extension of the contactless interface, so neither the mole nor the card emulator need to ``understand'' the actual communication. They simply proxy any bits of data they receive.

As existing cryptographic protocols on the application layer cannot prevent relay attacks, several alternative methods have been identified to prevent or hinder relay attacks\cite{Hancke2005,Kfir2005,Hancke2009}:
\begin{enumerate}
\item The card's radio frequency interface can be shielded with a Faraday cage (e.g.\ aluminium foil) when not in use.
\item The card could contain additional circuitry for physical activation and deactivation.
\item Additional passwords or PIN codes could be used for two-factor authentication.
\item Distance bounding protocols can be used on fast channels to determine the actual distance between the card and the reader.
\end{enumerate}

Other measures -- like measurement of command delays to detect additional delays induced by relay channels -- have been identified as not useful. For instance, Hancke et al.\cite{Hancke2009} conclude that the timing constraints of ISO/IEC 14443 are too loose to provide adequate protection against relay attacks.

\subsection{Next Generation Relay Attack}
The threat potential of relay attacks was mitigated by the fact that all relay scenarios required physical proximity (less than one meter) to the device under attack. However, recent research\cite{Roland2012,Roland2012IFIP} follows a different approach. Instead of accessing a device's secure element through the external (contactless) interface, it is accessed from the device's application processor through the internal interface. While the original relay attack required mole hardware in physical proximity of the device under attack, pure software on an attacked device's application processor is enough.

\begin{figure}[!ht]
  \centering
  \includegraphics[width=110mm]{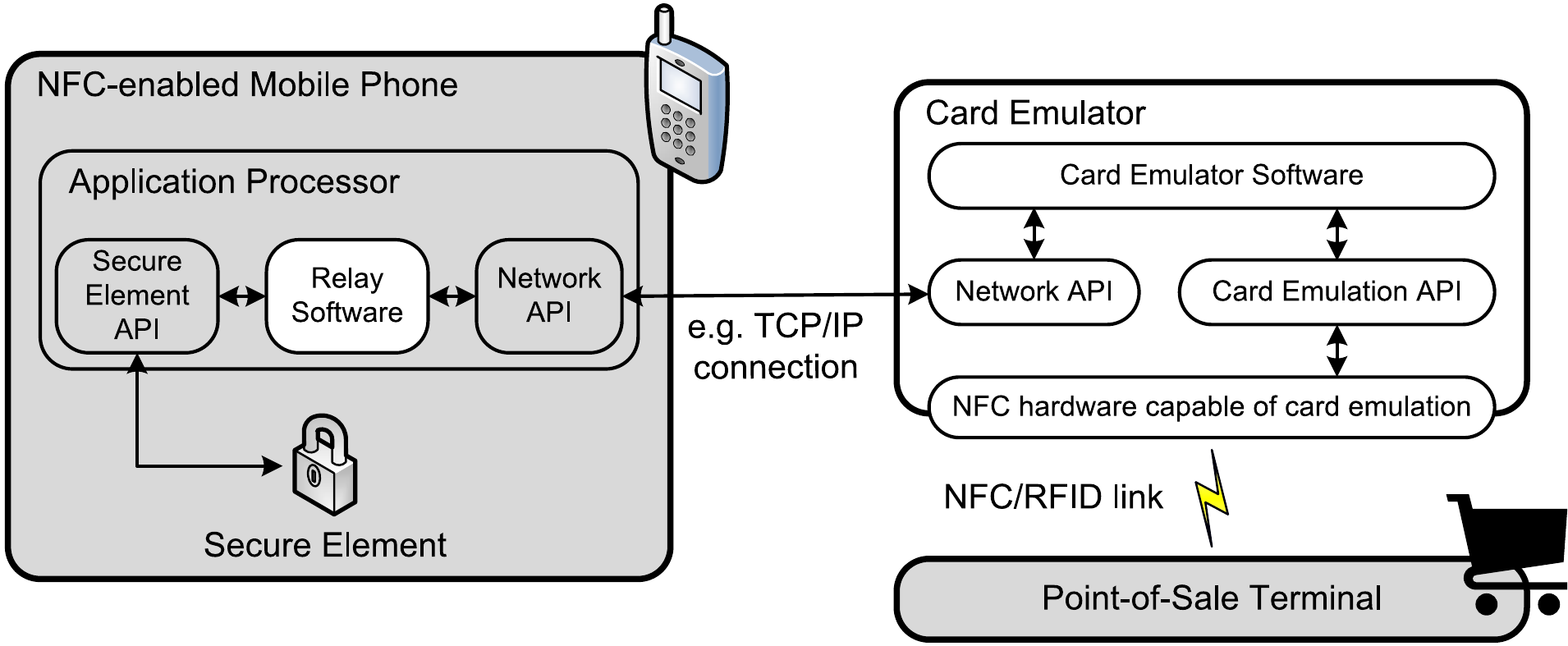}
  \caption{Relay scenario: Relay software is installed on the victim's phone. The software relays APDUs between the secure element and the card emulator across a network (cellular network, WiFi, Bluetooth...) The card emulator emulates a contactless smartcard that interacts with a card reader (point-of-sale terminal, access control reader...) The card emulator routes all APDU commands received from the point-of-sale terminal through the network interface to the relay software on the victim's mobile phone. As soon as the response APDU is received from the relay software, it is forwarded to the reader.}\label{fig:relaysystem}
\end{figure}

The complete relay system, as suggested by\cite{Roland2012} and verified in\cite{Roland2012IFIP} is shown in Fig.~\ref{fig:relaysystem}. It consists of four parts:
\begin{itemize}
	\item a mobile phone (under control of its owner/legitimate user),
	\item a relay software (under control of the attacker),
	\item a card emulator (under control of the attacker), and
	\item a reader device (e.g.\ at a point-of-sale terminal or at an access control gate).
\end{itemize}

The relay software is installed on the victim's mobile phone. This application is assumed to have the privileges necessary for access to the secure element and for communicating over a network. These privileges can be either explicitly granted to the application or acquired by means of a privilege escalation attack. The relay application waits for APDU commands on a network socket and forwards these APDUs to the secure element. The responses are then sent back through the network socket.

The card emulator is a device that is capable of emulating a contactless smartcard in software. The emulator has RFID/NFC hardware that acts as a contactless smartcard when put in front of a smartcard reader. The emulator software forwards the APDU commands (and responses) between a network socket and the emulator's RFID/NFC hardware.

\begin{figure}[!ht]
  \centering
  \includegraphics[width=115mm]{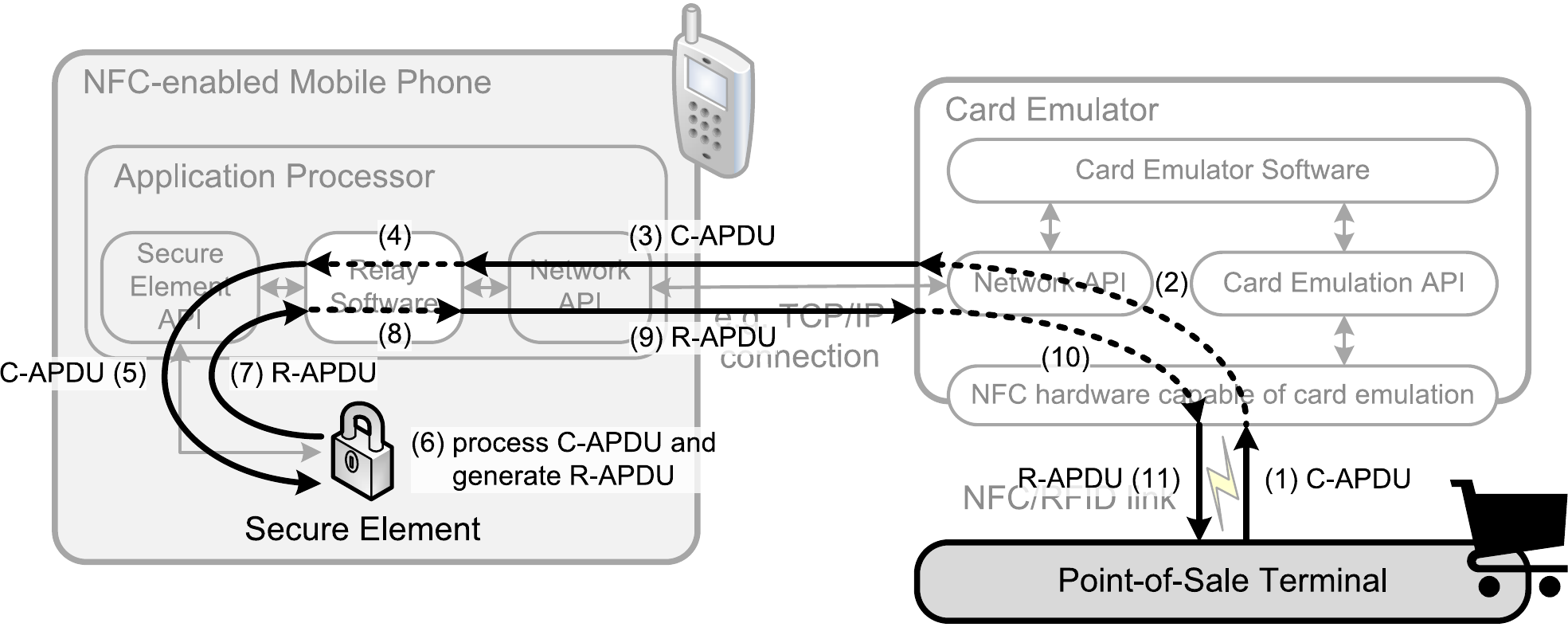}
  \caption{Flow of relayed smartcard commands (APDUs) between a smartcard reader and a secure element. The command APDUs (C-APDUs) received from the point-of-sale terminal are routed through the card emulator and over a wireless network to the victims device. There, the relay app forwards the C-APDUs to the secure element. The corresponding responses (R-APDUs) generated by the secure element are routed all the way back (through the relay app, the wireless network and the card emulator) to the POS terminal.}\label{fig:relayapdus}
\end{figure}
The flow of relayed smartcard commands (APDUs) between the smartcard reader and the secure element is shown in Fig.~\ref{fig:relayapdus}. The command APDUs (C-APDUs) received from the point-of-sale terminal are routed through the card emulator and over a wireless network to the victims device. There, the relay app forwards the C-APDUs to the secure element. The corresponding responses (R-APDUs) generated by the secure element are routed all the way back (through the relay app, the wireless network and the card emulator) to the POS terminal.

\subsection{Access to the Secure Element}
Various schemes for access control to the secure element are analyzed in\cite{Roland2012}. While some of them provide sophisticated access control capabilities, all of them have one significant flaw. They all rely on the mobile device's operating system (excuted on the application processor) to perform access control enforcement. Thus, in all cases, the secure element (secure component) blindly trusts the operating system's/application processor's (i.e.\ insecure component's) access control decisions. Therefore, once an application passes or bypasses(!) the security checks performed by the operating system, it can exchange (arbitrary) APDUs with the secure element.

Consequently, in the worst-case scenario, root access to the operating system is required to bypass these security checks. However, considering the current trend in privilege escalation exploits for various mobile device platforms (cf.\cite{Roland2012IFIP}), it is assumed that an arbitrary application can gain elevated or even root privileges on most platforms that are currently in the field.

For instance, for the Android platform, recent exploits comprise \emph{mempodroid} (Android 4.0 and later), \emph{Levitator} (up to Android 2.3.5), \emph{zergRush} (up to Android 2.3.3), \emph{GingerBreak} (up to Android 2.3.3), \emph{ZimperLich}, \emph{KillingInTheName}, \emph{RageAgainstTheCage}, \emph{Exploid} ...

\subsection{Suitable Relay Channels}
\begin{figure}[!ht]
  \centering
  \subfigure[Path~\ref{item:direct+external}]{%
            \includegraphics[width=.495\textwidth]{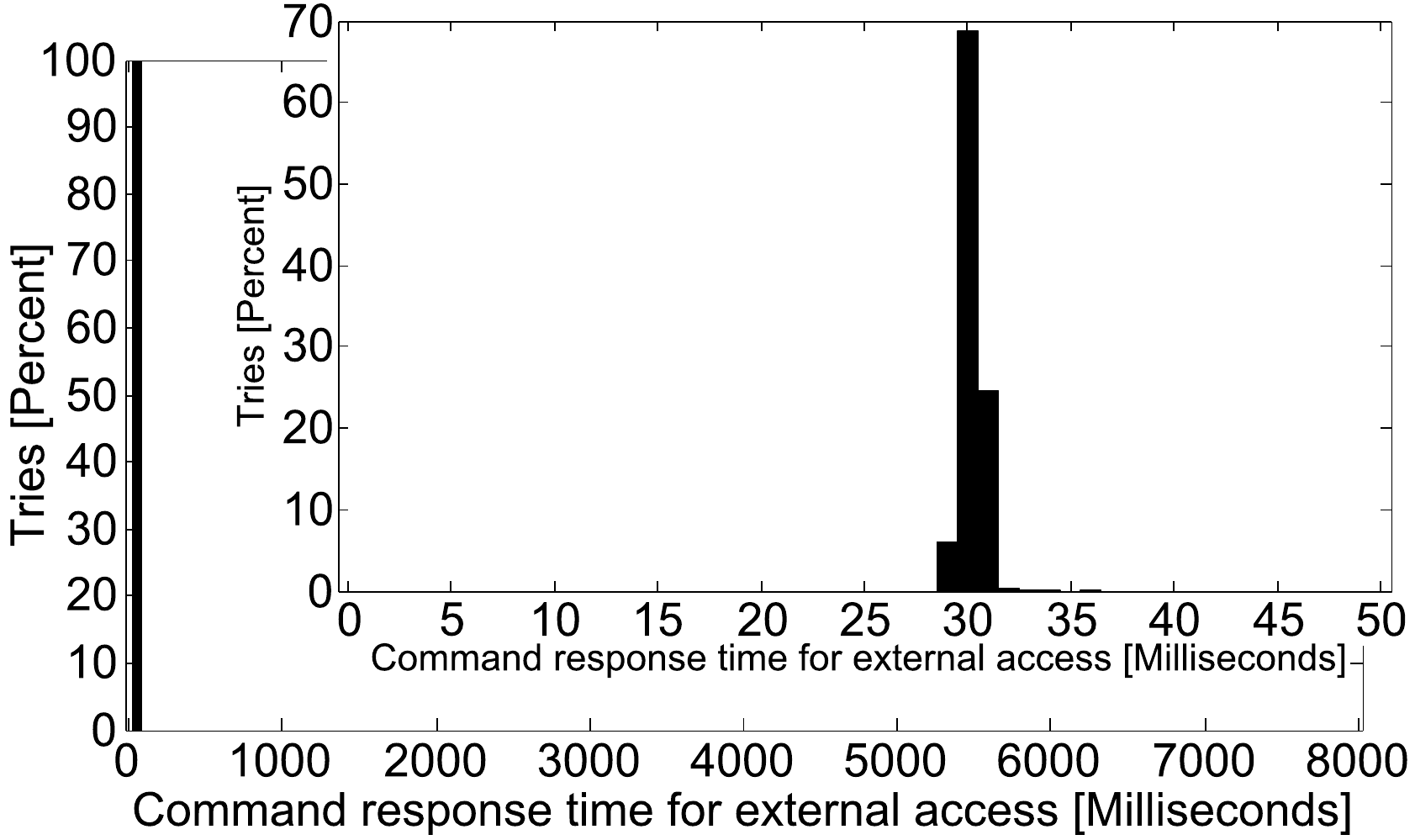}%
            \label{fig:histograms+selcm+ext}}
  \subfigure[Path~\ref{item:direct+internal}]{%
            \includegraphics[width=.495\textwidth]{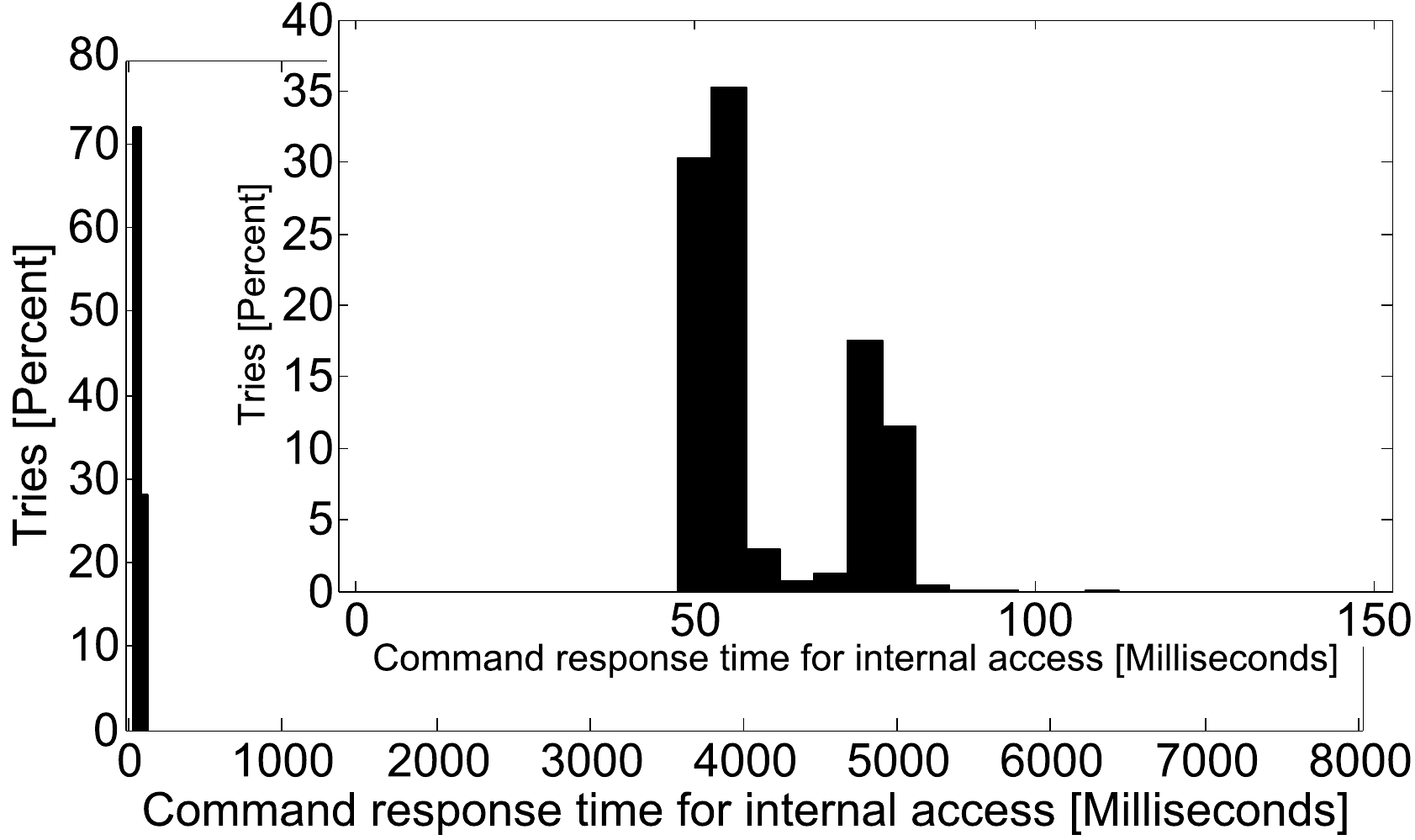}%
            \label{fig:histograms+selcm+int}}
  \subfigure[Path~\ref{item:relay+wifi}]{%
            \includegraphics[width=.495\textwidth]{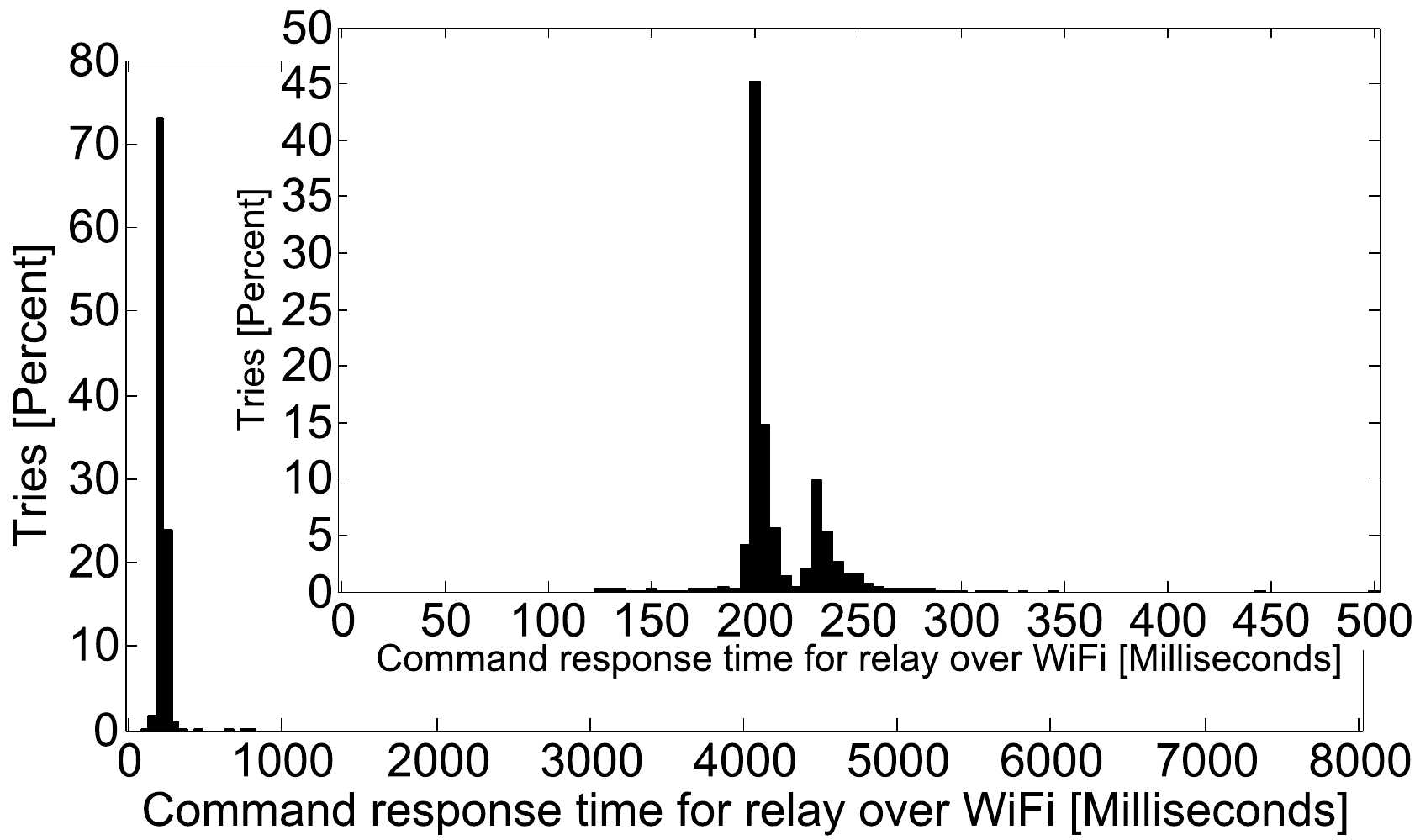}%
            \label{fig:histograms+selcm+relaywifi}}
  \subfigure[Path~\ref{item:relay+internet}]{%
            \includegraphics[width=.495\textwidth]{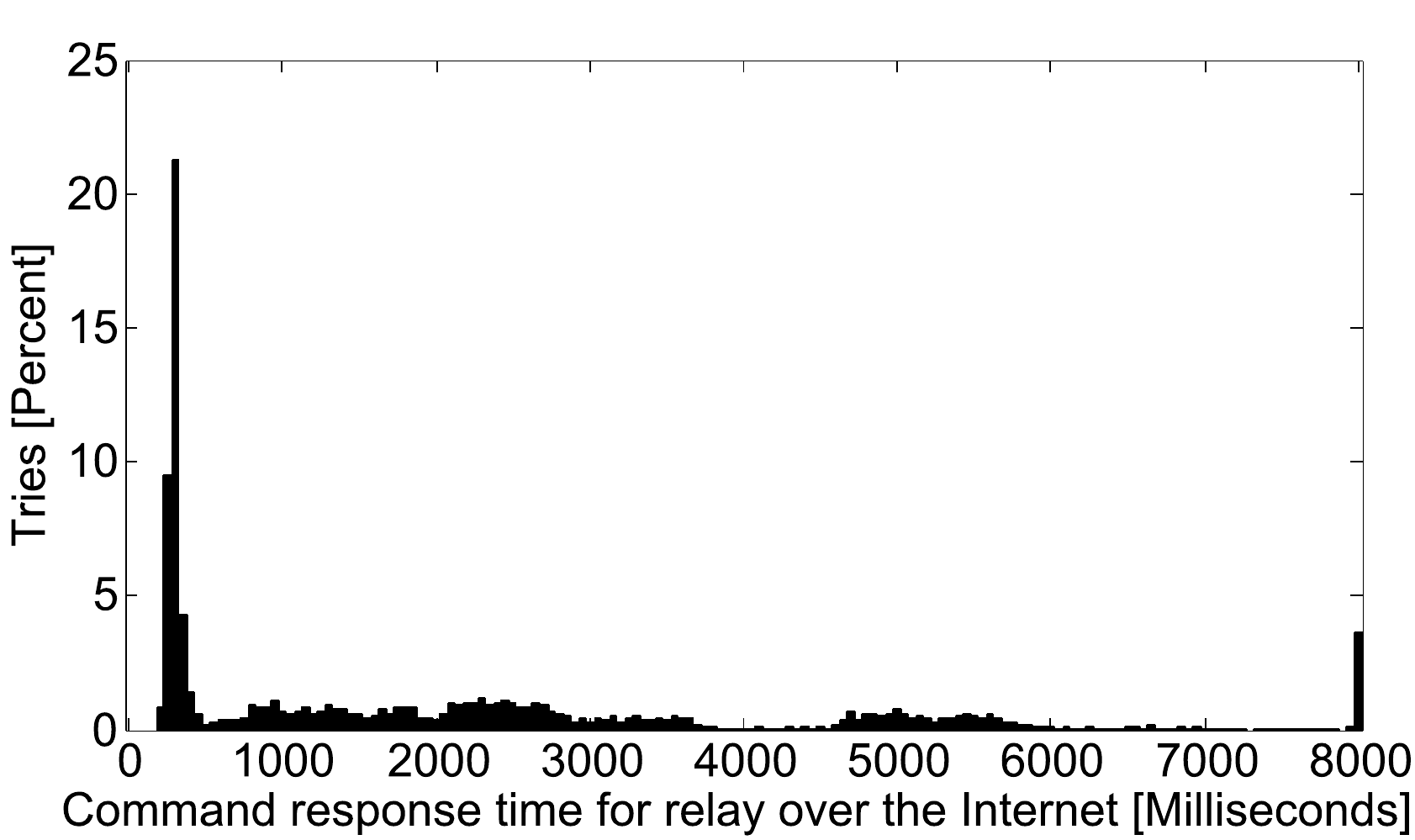}%
            \label{fig:histograms+selcm+relayinternet}}
  \caption{Histograms of delay between command and response at the reader side for the APDU ``SELECT Issuer Security Domain (card manager) by AID'' (C-APDU: 13 bytes, R-APDU: 105 bytes) for 5000 repetitions. The histogram is divided into 160 bins. Each bin has a width of $50\,\textrm{ms}$. The last bin also contains all measurements above $8000\,\textrm{ms}$. \subref{fig:histograms+selcm+ext} is zoomed from $0$ to $50\,\textrm{ms}$ with $1$-ms-bins. \subref{fig:histograms+selcm+int} is zoomed from $0$ to $150\,\textrm{ms}$ with $5$-ms-bins. \subref{fig:histograms+selcm+relaywifi} is zoomed from $0$ to $500\,\textrm{ms}$ with $5$-ms-bins.}%
  \label{fig:histograms+selcm}
\end{figure}
In\cite{Roland2012IFIP}, the delay times induced by relaying APDU communication over various channels are evaluated. Fig.~\ref{fig:histograms+selcm} shows a comparison of four different paths for secure element access:
\begin{enumerate}
	\item\label{item:direct+external} Direct access to the secure element with an external reader (i.e.\ no relay),
	\item\label{item:direct+internal} direct access to the secure element with an app on the phone,
	\item\label{item:relay+wifi} access through the relay system using a direct WiFi link between the phone and the card emulator,
	\item\label{item:relay+internet} access through the relay system using the mobile phone network and an Internet link between the phone and the card emulator.
\end{enumerate}
Direct access to the secure element through the contactless interface (path~\ref{item:direct+external}) takes about $30\,\textrm{ms}$. On-device access to the secure element (path~\ref{item:direct+internal}) takes significantly longer ($50$ to $80\,\textrm{ms}$). A WiFi connection (path~\ref{item:relay+wifi}) adds an additional delay in the range of $100$ and $210\,\textrm{ms}$. For path~\ref{item:relay+internet}, the delays start at about $200\,\textrm{ms}$ and have a significant peak around $300\,\textrm{ms}$. Thus, the internet link adds at least $150\,\textrm{ms}$ of delay. But for more than half of the measurements the total command-response delay was above $1$ second.

In practice there are no (strict) timing requirements for payment applications. ISO/IEC 14443-4 provides a \emph{frame waiting time extension} so that no timing requirements apply to the APDU layer. The EMV specification for contactless payment systems\cite{EMVContactless2011BookA} specifies a limit of $500\,\textrm{ms}$ for a contactless payment transaction as a whole (where a \emph{transaction} already comprises multiple APDU command-response sequences.) Consequently, both relay scenarios are likely to fail these timings. However, a payment terminal is not required to interrupt a transaction if it takes longer than this limit. The limit is merely meant as a benchmark target to maintain user experience. For example, the PayPass terminals used in recent roll-outs in Austria (see Fig.~\ref{fig:paypass+terminal+austria}) do not enforce any such timings. Also, cloud-based secure element solutions (cf.\cite{Roland2012IWSSI}) like those provided by YES-wallet\footnote{\url{http://www.yes-wallet.com/}} will only work with relaxed timing requirements.

\section{Applying the Attack to a Real-World Payment System}
\boxblock{Videos of the successful relay attack are available on YouTube:
\begin{itemize}
	\item Initial video proof:\\ \url{http://www.youtube.com/watch?v=hx5nbkDy6tc}
	\item Re-take, better quality:\\ \url{http://www.youtube.com/watch?v=_R2JVPJzufg}
\end{itemize}
}

To verify the applicability of the software-based relay attack, it has been applied to an existing payment system. \emph{Google Wallet} has been chosen for several reasons:
\begin{itemize}
	\item Google Wallet is already in use by many users. (Google Play Store listed more than 500,000 installations\footnote{\url{https://play.google.com/store/apps/details?id=com.google.android.apps.walletnfcrel}} in early 2012. Meanwhile Google Wallet has over 1,000,000 installations.)
	\item It is based on EMV payment standards (specifically \emph{MasterCard PayPass} using the EMV mag-stripe mode) and can be used with any point-of-sale terminal that supports PayPass contactless credit card transactions.
	\item The Android source code is publicly available. Thus, it was fairly easy to explore its NFC software stack and its hidden\footnote{\emph{Hidden} means that it is not included in the public software development kit.} secure element API (\texttt{com.{\nohyph}android.{\nohyph}nfc\_{\nohyph}extras}).
	\item Google Wallet is known to be installed by many users on rooted devices (mainly to circumvent operator and location restrictions). This means that the operating system's security measures are already weakened/bypassed on those phones.
	\item For non-rooted devices, there either already exist privilege escalation exploits (up to Android 2.3.5 and on Android 4.0+) or it is assumed that such exploits will appear soon (cf.\cite{Zvelo2012RootExploits}). Additionally, once an exploit is found/known, it takes several month until devices in the field are patched.
\end{itemize}

\subsection{Google Wallet}\label{sec:google+wallet}
Google Wallet is a container for payment cards, gift cards, reward cards and special offers. It consists of an Android app with a user interface and JavaCard applets on the secure element. The user interface is used to unlock the wallet (when it was previously locked by a PIN code), to select the currently active card, to find specific offers and to view the transaction history. The analysis and attack described in this paper have been performed with version 1.1-R52v7 of the Google Wallet app and the secure element applets installed in February 2012.

\boxblock{Google has quickly responded to our discoveries by providing fixes in more recent versions of Google Wallet. For instance, the relay attack could not be reproduced with secure element applets installed in June 2012. With recent upgrades of the Google Wallet app, also existing users received the necessary fixes of the secure element applets and are no longer vulnerable to the relay attack scenario described in this paper.}

Upon first start, the Google Wallet app initializes the secure element and installs a PIN code that is necessary for using the Google Wallet app's user interface. During initialization several applets are installed and personalized on the secure element using GlobalPlatform card management. Specifically, a secure channel based on the secure channel protocol SCP02 is established between the secure element and a remote server which performs the card management through this authenticated and (partly) encrypted channel.

Executable load-files on the secure element after initialization:
\begin{itemize}
	\item \texttt{A0000000035350}: Issuer Security Domain/Card Manager
	\item \texttt{A00000000410}: MasterCard Credit Card
	\item \texttt{A00000047610}: Unknown (Google)
	\item \texttt{A0000004761000}: Unknown (Google)
	\item \texttt{A0000004761001}: Unknown (Google)
	\item \texttt{A0000004761002}: Unknown (Google)
	\item \texttt{A00000047620}: Google Wallet
	\item \texttt{A00000047630}: Google Mifare Access
	\item \texttt{785041592E}: EMV Payment System Environment
\end{itemize}

Applets used by Google Wallet and payment terminals:
\begin{itemize}
	\item \texttt{A0000000041010}: MasterCard Credit Card
	\item \texttt{A0000000041010AA54303200FF01FFFF}: MasterCard Google Prepaid Card
	\item \texttt{A0000004762010}: Google Wallet On-Card Component
	\item \texttt{A0000004763030}: Google Mifare Access Applet
	\item \texttt{325041592E5359532E4444463031}: EMV Proximity Payment System Environment (PPSE, 2PAY.SYS.DDF01)
\end{itemize}
Google Wallet On-Card Component and Google Mifare Access Applet are only selectable through the secure element's internal mode but cannot be selected through the contactless interface.

Google Wallet APDU commands:
\begin{itemize}
	\item \texttt{00A4040007A000000476201000}: Select Google Wallet on-card component
	\item \texttt{80E200AA00}: Unlock Google Wallet (This command is used after successful PIN verification in the Google Wallet app. The PIN is not verified by the on-card component.)
	\item \texttt{80E2005500}: Lock Google Wallet
	\item \texttt{80CA00A500}: List installed payment cards(?)
	\item \texttt{80F24000024F0000}: Similar to GlobalPlatform GET STATUS for applications and supplementary security domains(?)
	\item \texttt{80F00101124F10A0000000041010AA54303200FF01FFFF00}: Disable Google Prepaid Card(?)
	\item \texttt{80F00201124F10A0000000041010AA54303200FF01FFFF00}: Enable Google Prepaid Card(?)
\end{itemize}
For a successful relay attack, it is necessary to select the Google Wallet on-card component and unlock the wallet. After this, the default payment card can be accessed through the secure element's internal mode. No user interaction is required. If other cards besides the MasterCard Google Prepaid Card are installed into the wallet, additional commands might be necessary to enable the desired payment card.

\subsection{Google Prepaid Card}
The Google Prepaid Card uses the \emph{EMV Contactless Specifications for Payment Systems} and is a MasterCard PayPass card. It only supports the Mag-Stripe mode with dynamic CVC3 (card verification code) and online transactions\footnote{Note that the attack described in this paper is expected to also work with EMV mode (also known as \emph{Chip \& PIN}), as long as cardholder verification (PIN code) is \emph{not} required for the transaction.}.

During a Mag-Stripe mode transaction, the following command-sequence is performed by a POS terminal:
\begin{enumerate}
	\item POS terminal: Select Proximity Payment System Environment (PPSE) (see Table~\ref{tab:select+ppse+c})
	\item Card: Confirm selection and respond with a list of supported EMV payment applications (see Table~\ref{tab:select+ppse+r})
	\item POS terminal: Select MasterCard Google Prepaid Card (see Table~\ref{tab:select+mcgpc+c})
	\item Card: Confirm selection and respond with application details (e.g.\ that it is a MasterCard) (see Table~\ref{tab:select+mcgpc+r})
	\item POS terminal: Get processing options (see Table~\ref{tab:gpo+c})
	\item Card: Respond with supported mode (Mag-Stripe only, online transactions only, no cardholder verification...) and with the location of the Mag-Stripe data file (see Table~\ref{tab:gpo+r})
	\item POS terminal: Read first record data file (see Table~\ref{tab:read+magstripe+c})
	\item Card: Return Mag-Stripe track 1 and track 2 data (see Table~\ref{tab:read+magstripe+r})
	\item POS terminal: Request computation of cryptographic checksum for a given unpredictable number (UN) (see Table~\ref{tab:compute+cc+c})
	\item Card: Return application transaction counter (ATC) and dynamically generated CVC3 for track 1 and track 2 (see Table~\ref{tab:compute+cc+r})
\end{enumerate}

\subsection{Android's Secure Element API}
Android's secure element API (\texttt{com.{\nohyph}android.{\nohyph}nfc\_{\nohyph}extras}) is available since Android 2.3.4. It consists of two classes: NfcAdapterExtras (see~\ref{appendix:nfcadapterextras}) and NfcExecutionEnvironment (see~\ref{appendix:nfcexecutionenvironment}). NfcAdapterExtras is used to enable and disable external card emulation and to retrieve an instance of the embedded secure element's NfcExecutionEnvironment class. NfcExecutionEnvironment is used to establish an internal connection to a secure element and to exchange APDUs with it.

In Android 2.3.4, this API could be accessed by any application that held the permission to use NFC. In later versions, a special permission named \texttt{com.{\nohyph}android.{\nohyph}nfc.{\nohyph}permission.{\nohyph}NFCEE\_{\nohyph}ADMIN} is required. This permission is only granted to applications that are signed with the same certificate as the NFC system service. Starting in Android 4.0, the permission system for the secure element API has fundamentally changed. Permissions to access the secure element are now granted through an XML file. This XML file contains a list of certificates that are granted access. However, applications with root access can easily obtain the permission to access the secure element for any of these access control mechanisms.

\subsection{The Relay App}
The relay app, a purely Java-based Android app, is a simple TCP client that maintains a persistent TCP connection to a remote server (the card emulator). When the card emulator requests access to the secure element, a connection is established through the NfcExecutionEnvironment object. Then, the Google Wallet on-card component is selected and the unlock command is sent. The relay app then listens for C-APDUs on its network interface and forwards them to the secure element. The R-APDUs from the secure element are transmitted back to the card emulator. When the transaction is complete, the Google Wallet on-card component is selected again and the lock command is sent to lock the wallet.

For this test scenario, the relay app has been manually granted the permissions necessary to access the secure element. However, privilege escalation exploits could be integrated into future versions of the app. For easier integration of future exploit codes, a privilege escalation framework (cf.\cite{Hobarth2011}) could be embedded into the app.

Additionally, the test app has a foreground component that needs to be started manually. Moreover, the connection to the card emulator needs to be confirmed by the user. However, the app could be started automatically on device boot-up and run completely in the background.

To roll out the relay app to user's devices, it could be integrated into any existing app downloaded from Google Play Store. The infected app could then be re-published on Google Play Store under similar (or even identical) publisher information and with the same app name as its original\footnote{Note that Google started to combat this with recent updates to the Google Play Developer Program Policy.}. To specifically target users of rooted devices, an app that already requires root permissions could be used as a base for code injection.

\subsection{The Card Emulator}
\begin{figure}[!ht]
  \centering
  \includegraphics[width=160mm]{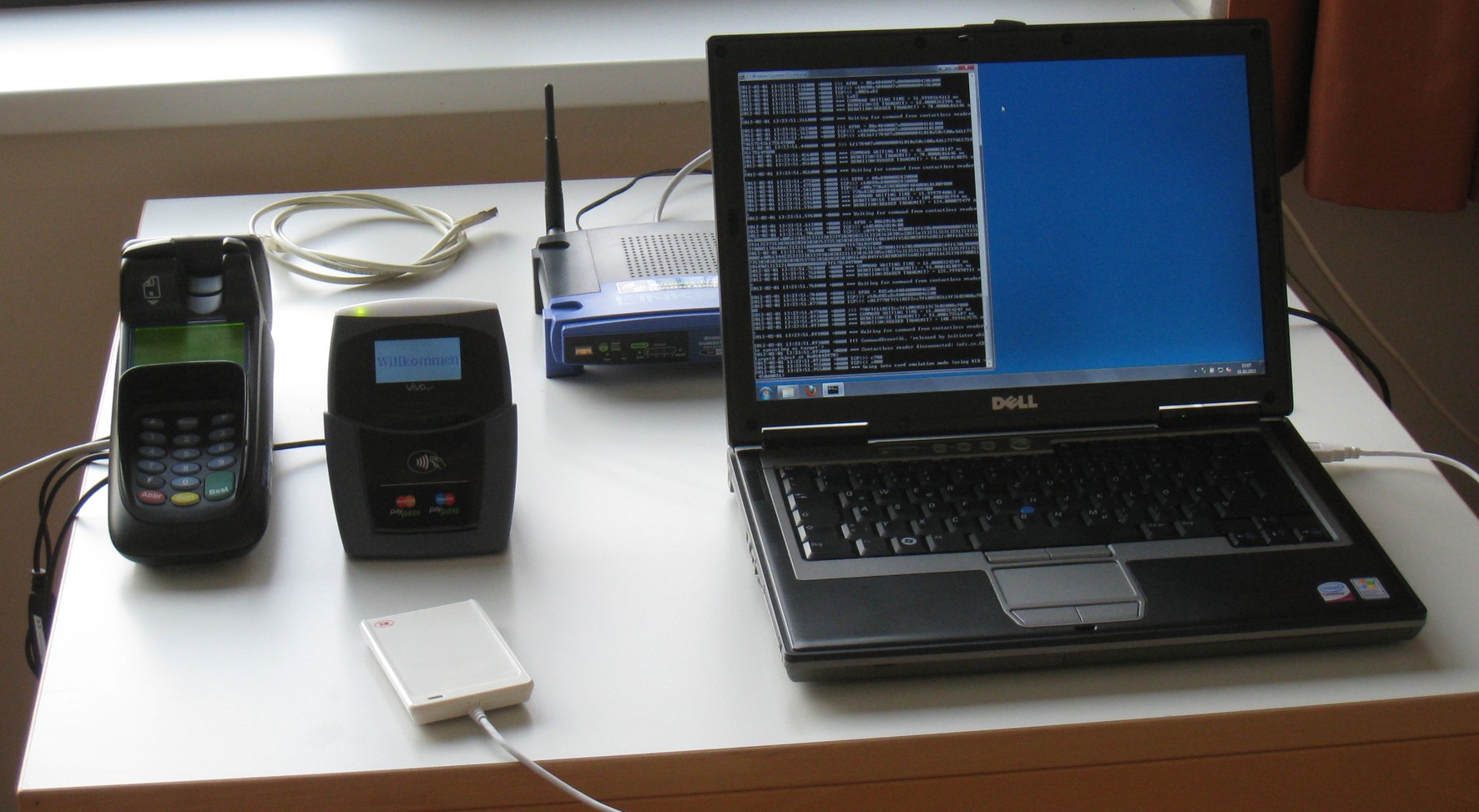}
  \caption{Card emulator made from a notebook and an ACS ACR 122U NFC reader.}\label{fig:custom+card+emulator}
\end{figure}

For this proof-of-concept a simple card emulator has been built from a notebook computer and an ACS ACR 122U NFC reader (Fig.~\ref{fig:custom+card+emulator}). This NFC reader supports software card emulation mode and is available for less than EUR 50 (including taxes and shipping) from touchatag.com\footnote{\url{http://store.touchatag.com/acatalog/touchatag_starter_pack.html}}. Several examples on how to use this device in card emulation mode can be found on the web\footnote{E.g.\ on \url{http://www.libnfc.org/}}.

The card emulation software (written in Python) contains a TCP server that listens for incoming connections from the relay app. Once a TCP connection has been established, the emulator goes into card emulation mode and waits for a POS terminal. When the card emulator detects activation by a POS terminal (or any smartcard reader), it requests access to the secure element through the relay app. Then, all received C-APDUs are forwarded through the network interface to the relay app and all R-APDUs received from the relay app are returned to the POS terminal. When the RF field is deactivated, the connection to the secure element is closed.

\begin{figure}[!ht]
  \centering
  \includegraphics[width=40mm]{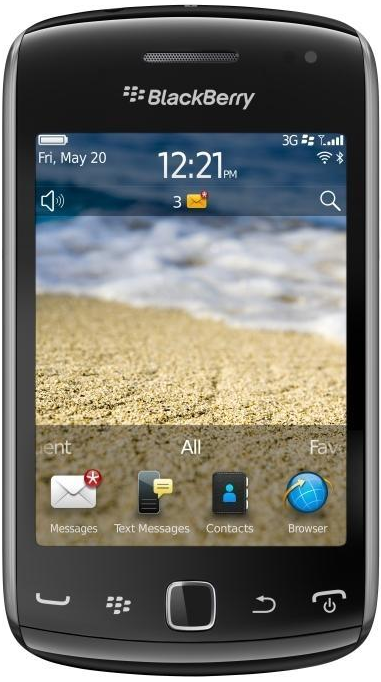}
  \caption{BlackBerry 9380 supports software card emulation. (Source:\\ \href{http://www.phonearena.com/news/BlackBerry-Curve-9380-announced-the-first-ever-Curve-with-a-touchscreen_id23783}{http://www.phonearena.com/news/BlackBerry-Curve-9380-announced-the-first-ever-Curve-with-a-touchscreen\_id23783})}\label{fig:blackberry9380}
\end{figure}

The custom card emulator in Fig.~\ref{fig:custom+card+emulator} has a form factor that will certainly raise suspicions at the point-of-sale. However, there already exist devices with an accepted form factor (i.e.\ that of a mobile phone, see Fig.~\ref{fig:blackberry9380}) with support for software card emulation. E.g.\ all BlackBerry devices with NFC that are equipped with the BlackBerry 7 platform support software card emulation mode. Moreover, recent patches\cite{cmPatchSoftwareCE1,cmPatchSoftwareCE2} to the CyanogenMod firmware for Android devices enable software card emulation for Android NFC devices that are based on the PN544 NFC controller. Besides their form factor, mobile phones have anonther advantage when used as card emulator platform for relay attacks: They already contain the same network interfaces as the device under attack\cite{Roland2012IWSSI}. The viability of a BlackBerry device as card emulator platform for relay attacks has already been verified by Francis et al.\cite{Francis2011}.

\subsection{Point-of-Sale (POS) Terminal}
\begin{figure}[!ht]
  \centering
  \includegraphics[width=80mm]{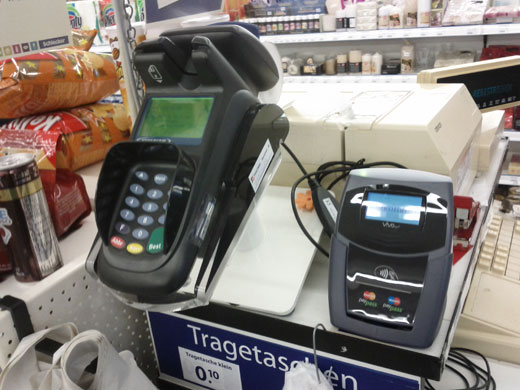}
  \caption{POS terminal (HYPERCOM Artema Hybrid) with contactless module (Vi\-VO\-tech 5000), as used in recent roll-outs at Schlecker and Zielpunkt in Austria. (Source: nfc.cc\protect\cite{PayPassLaunchNFCCC2011})}\label{fig:paypass+terminal+austria}
\end{figure}

The Google Wallet relay attack has been successfully tested together with a POS terminal as used in recent roll-outs at Schlecker and Zielpunkt in Austria (Fig.~\ref{fig:paypass+terminal+austria}).
\begin{figure}[!ht]
  \centering
  \includegraphics[width=93mm]{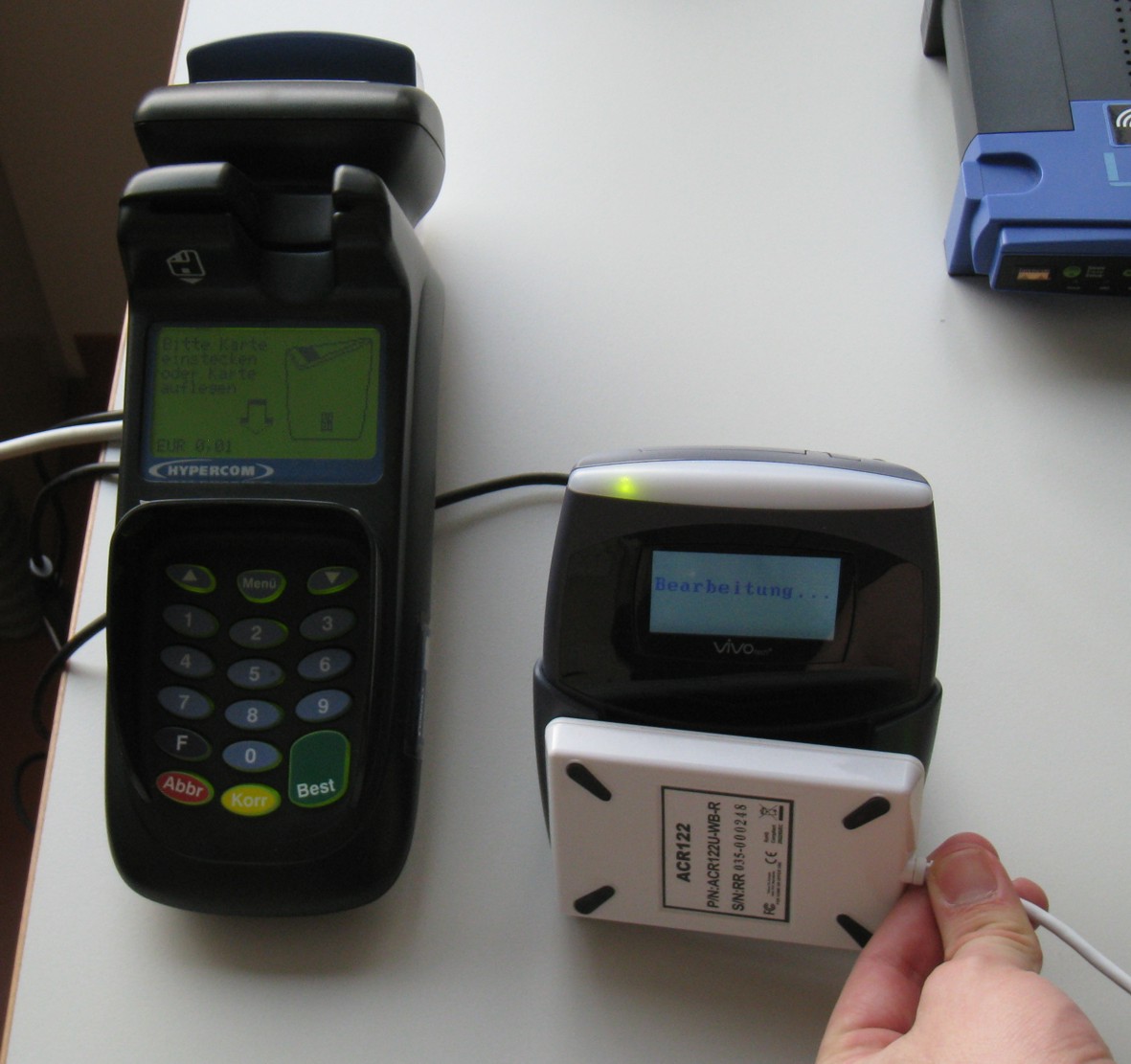}
  \caption{Successful payment transaction using the Google Wallet relay attack.}\label{fig:wallet+relay+attack+payment}
\end{figure}

\subsection{Viability of the Google Wallet Relay Attack}
For this attack only an NFC reader device (available for less than EUR 50), a notebook computer and some average programming skills were necessary. A BlackBerry with software card emulation support is available for less than EUR 300. An additional EUR 20 is necessary for a publisher account on Google Play Store.

When the relay app is running on many devices, a ``bot network'' of Google Wallets could be created. The attacker could then perform some kind of ``load balancing'' to evenly distribute payments among devices and to select a device with a stable network connection for each payment transaction.

\section{Possible Workarounds}

\subsection{Timeouts of POS Terminals}
An easy, but potentially unreliable, measure to prevent relay attacks would be the enforcement of short timeouts (e.g.\ those specified by the EMV specifications) for payment transactions on the POS terminals. Transactions taking longer than this timeout should be interupted or discarded. While this measure will prevent most long-distance relay scenarios, relays over shorter distances and fast communication channels might not be rejected. Also, installing such tight timeouts will prevent cloud-based EMV applications (cf.\cite{Roland2012IWSSI} and YES-wallet\footnote{\url{http://www.yes-wallet.com/}}).

\subsection{Google Wallet PIN Code Verification}
With the tested implementation of the Google Wallet app (version 1.1-R52v7), the PIN code that protects the Google Wallet is only verified within the mobile phone app. The on-card component does not verify this PIN code. Instead, the on-card component is controlled by simple lock and unlock commands (see \ref{sec:google+wallet}).

PIN code verification could be handled by the on-card component on the secure element instead of by the app on the mobile phone's application processor. After all, PIN code verification is a core component of smartcards anyways. In that case, the attacker would need to know the wallet's PIN code to conduct a successful attack.

On a rooted device, an attacker might still be able to sniff the PIN code (while it is entered by the user) by capturing the screen and the touch events. However, this is significantly more difficult than sending a simple unlock command to the secure element.

\subsection{Disabling Internal Mode Communication for Payment Applets}
Recent secure elements (like the one embedded into the Nexus S) provide instruments to distinguish between external communication (contactless interface) and internal communication (application processor) from within a JavaCard applet. In addition, the secure element may have the capability to completely disable a certain interface on a per-applet basis. These capabilities could be used to disable internal mode communication for all payment applets and consequently disable their vulnerability for software-based relay attacks.

The disadvantage of this workaround is that the secure element cannot be used for future on-device secure payment applications (e.g.\ EMV payment in the mobile phone's webbrowser). Such applications would, however, be one of the major benefits of having a secure element inside a mobile phone.

\boxblock{Note: This method has been used to circumvent relay attacks in more recent versions of Google Wallet.}

\clearpage\phantomsection\addcontentsline{toc}{section}{\refname}
\bibliographystyle{gerabbrv2}
\bibliography{literature}

\clearpage
\startappendix
\section{APDUs for Mag-Stripe Mode Transactions}
\ctable[%
caption={Select Proximity Payment System Environment (PPSE): C-APDU},%
captionskip=-3pt,
label={tab:select+ppse+c},%
pos={!ht},%
center,%
mincapwidth={\hsize},%
doinside={\footnotesize},%
width={\textwidth},%
]{p{9mm}p{26mm}llllll}{%
}{%
\toprule[1pt]
CLA  & \ttfamily 00 & \multicolumn{6}{l}{Inter-industry class} \\\midrule[0.5pt]%
INS  & \ttfamily A4 & \multicolumn{6}{l}{Select} \\\midrule[0.5pt]%
P1   & \ttfamily 04 & \multicolumn{6}{l}{Selection by DF name} \\\midrule[0.5pt]%
P2   & \ttfamily 00 & \multicolumn{6}{l}{Return FCI template} \\\midrule[0.5pt]%
Lc   & \ttfamily 0E & \multicolumn{6}{l}{} \\\midrule[0.5pt]%
DATA & \ttfamily 32 50 41 59 2E 53 59 53 2E 44 44 46 30 31 & \multicolumn{6}{l}{Proximity Payment System Environment (2PAY.SYS.DDF01)} \\\midrule[0.5pt]%
Le   & \ttfamily 00 & \multicolumn{6}{l}{} \\\bottomrule[1pt]\addlinespace[-9pt]%
}%
\ctable[%
caption={Select Proximity Payment System Environment (PPSE): R-APDU},%
captionskip=-3pt,
label={tab:select+ppse+r},%
pos={!ht},%
center,%
mincapwidth={\hsize},%
doinside={\footnotesize},%
width={\textwidth},%
]{p{9mm}p{26mm}llllll}{%
}{%
\toprule[1pt]
DATA & \ttfamily 6F 3A 84 0E 32 & \multicolumn{5}{l}{Tag: FCI template (\texttt{6F})} \\%
     & \ttfamily 50 41 59 2E 53 & \multicolumn{5}{l}{Length: 58 (\texttt{3A})} \\%
     & \ttfamily 59 53 2E 44 44 & \multicolumn{5}{l}{Value: (constructed)} \\%
     & \ttfamily 46 30 31 A5 28 & & & \multicolumn{4}{l}{Tag: DF name (\texttt{84})} \\%
     & \ttfamily BF 0C 25 61 15 & & & \multicolumn{4}{l}{Length: 14 (\texttt{0E})} \\%
     & \ttfamily 4F 10 A0 00 00 & & & \multicolumn{4}{l}{Value: 2PAY.SYS.DDF01 (\texttt{325041592E5359532E4444463031})} \\%
     & \ttfamily 00 04 10 10 AA & & & \multicolumn{4}{l}{\ \ \ \ \ \ \ \ \ Proximity Payment System Environment} \\%
     & \ttfamily 54 30 32 00 FF & & & \multicolumn{4}{l}{Tag: Proprietary information encoded in BER-TLV (\texttt{A5})} \\%
     & \ttfamily 01 FF FF 87 01 & & & \multicolumn{4}{l}{Length: 40 (\texttt{28})} \\%
     & \ttfamily 01 61 0C 4F 07 & & & \multicolumn{4}{l}{Value: (constructed)} \\%
     & \ttfamily A0 00 00 00 04 & & & & \multicolumn{3}{l}{Tag: FCI issuer discetionary data (\texttt{BF0C})} \\%
     & \ttfamily 10 10 87 01 02 & & & & \multicolumn{3}{l}{Length: 37 (\texttt{25})} \\%
     & \ttfamily                & & & & \multicolumn{3}{l}{Value: (constructed)} \\%
     & \ttfamily                & & & & & \multicolumn{2}{l}{Tag: Application template (\texttt{61})} \\%
     & \ttfamily                & & & & & \multicolumn{2}{l}{Length: 21 (\texttt{15})} \\%
     & \ttfamily                & & & & & \multicolumn{2}{l}{Value: (constructed)} \\%
     & \ttfamily                & & & & & & Tag: Application identifier (\texttt{4F}) \\%
     & \ttfamily                & & & & & & Length: 16 (\texttt{10}) \\%
     & \ttfamily                & & & & & & Value: MasterCard Google Prepaid Card \\%
     & \ttfamily                & & & & & & \ \ \ \ \ \ \ \ \ (\texttt{A0000000041010AA54303200FF01FFFF}) \\%
     & \ttfamily                & & & & & & Tag: Application priority indicator (\texttt{87}) \\%
     & \ttfamily                & & & & & & Length: 1 (\texttt{01}) \\%
     & \ttfamily                & & & & & & Value: 1 (\texttt{01})\\%
     & \ttfamily                & & & & & \multicolumn{2}{l}{Tag: Application template (\texttt{61})} \\%
     & \ttfamily                & & & & & \multicolumn{2}{l}{Length: 12 (\texttt{0C})} \\%
     & \ttfamily                & & & & & \multicolumn{2}{l}{Value: (constructed)} \\%
     & \ttfamily                & & & & & & Tag: Application identifier (\texttt{4F}) \\%
     & \ttfamily                & & & & & & Length: 7 (\texttt{07}) \\%
     & \ttfamily                & & & & & & Value: MasterCard credit/debit card \\%
     & \ttfamily                & & & & & & \ \ \ \ \ \ \ \ \ (\texttt{A0000000041010}) \\%
     & \ttfamily                & & & & & & Tag: Application priority indicator (\texttt{87}) \\%
     & \ttfamily                & & & & & & Length: 1 (\texttt{01}) \\%
     & \ttfamily                & & & & & & Value: 2 (\texttt{02})\\\midrule[0.5pt]%
SW1  & \ttfamily 90 & \multicolumn{6}{l}{Success} \\%
SW2  & \ttfamily 00 & \multicolumn{6}{l}{} \\\bottomrule[1pt]\addlinespace[-9pt]%
}%

\ctable[%
caption={Select MasterCard Google Prepaid Card: C-APDU},%
captionskip=-3pt,
label={tab:select+mcgpc+c},%
pos={!ht},%
center,%
mincapwidth={\hsize},%
doinside={\footnotesize},%
width={\textwidth},%
]{p{9mm}p{26mm}llllll}{%
}{%
\toprule[1pt]
CLA  & \ttfamily 00 & \multicolumn{6}{l}{Inter-industry class} \\\midrule[0.5pt]%
INS  & \ttfamily A4 & \multicolumn{6}{l}{Select} \\\midrule[0.5pt]%
P1   & \ttfamily 04 & \multicolumn{6}{l}{Selection by DF name} \\\midrule[0.5pt]%
P2   & \ttfamily 00 & \multicolumn{6}{l}{Return FCI template} \\\midrule[0.5pt]%
Lc   & \ttfamily 10 & \multicolumn{6}{l}{} \\\midrule[0.5pt]%
DATA & \ttfamily A0 00 00 00 04 10 10 AA 54 30 32 00 FF 01 FF FF & \multicolumn{6}{l}{MasterCard Google Prepaid Card} \\\midrule[0.5pt]%
Le   & \ttfamily 00 & \multicolumn{6}{l}{} \\\bottomrule[1pt]\addlinespace[-9pt]%
}%
\ctable[%
caption={Select MasterCard Google Prepaid Card: R-APDU},%
captionskip=-3pt,
label={tab:select+mcgpc+r},%
pos={!ht},%
center,%
mincapwidth={\hsize},%
doinside={\footnotesize},%
width={\textwidth},%
]{p{9mm}p{26mm}llllll}{%
}{%
\toprule[1pt]
DATA & \ttfamily 6F 20 84 10 A0 & \multicolumn{5}{l}{Tag: FCI template (\texttt{6F})} \\%
     & \ttfamily 00 00 00 04 10 & \multicolumn{5}{l}{Length: 32 (\texttt{20})} \\%
     & \ttfamily 10 AA 54 30 32 & \multicolumn{5}{l}{Value: (constructed)} \\%
     & \ttfamily 00 FF 01 FF FF & & & \multicolumn{4}{l}{Tag: DF name (\texttt{84})} \\%
     & \ttfamily A5 0C 50 0A 4D & & & \multicolumn{4}{l}{Length: 16 (\texttt{10})} \\%
     & \ttfamily 61 73 74 65 72 & & & \multicolumn{4}{l}{Value: MasterCard Google Prepaid Card} \\%
     & \ttfamily 43 61 72 64    & & & \multicolumn{4}{l}{\ \ \ \ \ \ \ \ \ (\texttt{A0000000041010AA54303200FF01FFFF})} \\%
     & \ttfamily                & & & \multicolumn{4}{l}{Tag: Proprietary information encoded in BER-TLV (\texttt{A5})} \\%
     & \ttfamily                & & & \multicolumn{4}{l}{Length: 12 (\texttt{0C})} \\%
     & \ttfamily                & & & \multicolumn{4}{l}{Value: (constructed)} \\%
     & \ttfamily                & & & & \multicolumn{3}{l}{Tag: Application label (\texttt{50})} \\%
     & \ttfamily                & & & & \multicolumn{3}{l}{Length: 10 (\texttt{0A})} \\%
     & \ttfamily                & & & & \multicolumn{3}{l}{Value: MasterCard (\texttt{4D617374657243617264})} \\\midrule[0.5pt]%
SW1  & \ttfamily 90 & \multicolumn{6}{l}{Success} \\%
SW2  & \ttfamily 00 & \multicolumn{6}{l}{} \\\bottomrule[1pt]\addlinespace[-9pt]%
}%

\ctable[%
caption={Get Processing Options: C-APDU},%
captionskip=-3pt,
label={tab:gpo+c},%
pos={!ht},%
center,%
mincapwidth={\hsize},%
doinside={\footnotesize},%
width={\textwidth},%
]{p{9mm}p{26mm}llllll}{%
}{%
\toprule[1pt]
CLA  & \ttfamily 80 & \multicolumn{6}{l}{Proprietary class} \\\midrule[0.5pt]%
INS  & \ttfamily A8 & \multicolumn{6}{l}{Get Processing Options} \\\midrule[0.5pt]%
P1   & \ttfamily 00 & \multicolumn{6}{l}{} \\\midrule[0.5pt]%
P2   & \ttfamily 00 & \multicolumn{6}{l}{} \\\midrule[0.5pt]%
Lc   & \ttfamily 02 & \multicolumn{6}{l}{} \\\midrule[0.5pt]%
DATA & \ttfamily 83 00 & \multicolumn{6}{l}{Processing options data object list (PDOL) related data} \\%
     & \ttfamily       & \multicolumn{6}{l}{Tag: Command template (\texttt{83})} \\%
     & \ttfamily       & \multicolumn{6}{l}{Length: 0 (\texttt{00})} \\%
     & \ttfamily       & \multicolumn{6}{l}{Value: (empty)} \\\midrule[0.5pt]%
Le   & \ttfamily 00 & \multicolumn{6}{l}{} \\\bottomrule[1pt]\addlinespace[-9pt]%
}%
\ctable[%
caption={Get Processing Options: R-APDU},%
captionskip=-3pt,
label={tab:gpo+r},%
pos={!ht},%
center,%
mincapwidth={\hsize},%
doinside={\footnotesize},%
width={\textwidth},%
]{p{9mm}p{26mm}llllll}{%
}{%
\toprule[1pt]
DATA & \ttfamily 77 0A 82 02 00 & \multicolumn{5}{l}{Tag: Response message template (\texttt{77})} \\%
     & \ttfamily 00 94 04 08 01 & \multicolumn{5}{l}{Length: 10 (\texttt{0A})} \\%
     & \ttfamily 01 00          & \multicolumn{5}{l}{Value: (constructed)} \\%
     & \ttfamily                & & & \multicolumn{4}{l}{Tag: Application interchange profile (\texttt{82})} \\%
     & \ttfamily                & & & \multicolumn{4}{l}{Length: 2 (\texttt{02})} \\%
     & \ttfamily                & & & \multicolumn{4}{l}{Value: \texttt{00 00}} \\%
     & \ttfamily                & & & \multicolumn{4}{l}{\ \ \ \ \ \ \ \ \ Bit 1.7 = 0: no offline static data authentication supported} \\%
     & \ttfamily                & & & \multicolumn{4}{l}{\ \ \ \ \ \ \ \ \ Bit 1.6 = 0: no standard offline dynamic data authentication} \\%
     & \ttfamily                & & & \multicolumn{4}{l}{\ \ \ \ \ \ \ \ \ \ \ \ \ \ \ \ \ \ \ \ \ \ \ \ \ \ supported} \\%
     & \ttfamily                & & & \multicolumn{4}{l}{\ \ \ \ \ \ \ \ \ Bit 1.5 = 0: no cardholder verification supported} \\%
     & \ttfamily                & & & \multicolumn{4}{l}{\ \ \ \ \ \ \ \ \ Bit 1.4 = 0: no terminal risk management is to be performed} \\%
     & \ttfamily                & & & \multicolumn{4}{l}{\ \ \ \ \ \ \ \ \ Bit 1.3 = 0: no issuer authentication supported} \\%
     & \ttfamily                & & & \multicolumn{4}{l}{\ \ \ \ \ \ \ \ \ Bit 1.2 = 0: no combined DDA/AC generation supported} \\%
     & \ttfamily                & & & \multicolumn{4}{l}{\ \ \ \ \ \ \ \ \ Bit 2.8 = 0: only Mag-Stripe profile supported} \\%
     & \ttfamily                & & & \multicolumn{4}{l}{\ \ \ \ \ \ \ \ \ others: RFU} \\%
     & \ttfamily                & & & \multicolumn{4}{l}{Tag: Application file locator (\texttt{94})} \\%
     & \ttfamily                & & & \multicolumn{4}{l}{Length: 4 (\texttt{04})} \\%
     & \ttfamily                & & & \multicolumn{4}{l}{Value: \texttt{08 01 01 00}} \\%
     & \ttfamily                & & & \multicolumn{4}{l}{\ \ \ \ \ \ \ \ \ Bit 1.8-1.4 = 00001: Short EF = 1} \\%
     & \ttfamily                & & & \multicolumn{4}{l}{\ \ \ \ \ \ \ \ \ Bit 1.3-1.1 = 000} \\%
     & \ttfamily                & & & \multicolumn{4}{l}{\ \ \ \ \ \ \ \ \ Bit 2.8-2.1 = 00000001: First record to read is 1} \\%
     & \ttfamily                & & & \multicolumn{4}{l}{\ \ \ \ \ \ \ \ \ Bit 3.8-3.1 = 00000001: Last record to read is 1} \\%
     & \ttfamily                & & & \multicolumn{4}{l}{\ \ \ \ \ \ \ \ \ Bit 4.8-4.1 = 00000000: 0 consecutive records signed in} \\%
     & \ttfamily                & & & \multicolumn{4}{l}{\ \ \ \ \ \ \ \ \ \ \ \ \ \ \ \ \ \ \ \ \ \ \ \ \ \ \ \ \ \ \ \ \ \ \ \ \ \ \ \ \ Signed Application} \\\midrule[0.5pt]%
SW1  & \ttfamily 90 & \multicolumn{6}{l}{Success} \\%
SW2  & \ttfamily 00 & \multicolumn{6}{l}{} \\\bottomrule[1pt]\addlinespace[-9pt]%
}%

\ctable[%
caption={Read first record data file: C-APDU},%
captionskip=-3pt,
label={tab:read+magstripe+c},%
pos={!ht},%
center,%
mincapwidth={\hsize},%
doinside={\footnotesize},%
width={\textwidth},%
]{p{9mm}p{26mm}llllll}{%
}{%
\toprule[1pt]
CLA  & \ttfamily 00 & \multicolumn{6}{l}{Inter-industry class} \\\midrule[0.5pt]%
INS  & \ttfamily B2 & \multicolumn{6}{l}{Read record(s)} \\\midrule[0.5pt]%
P1   & \ttfamily 01 & \multicolumn{6}{l}{Record number 1} \\\midrule[0.5pt]%
P2   & \ttfamily 0C & \multicolumn{6}{l}{Short EF = 1, Read record P1} \\\midrule[0.5pt]%
Le   & \ttfamily 00 & \multicolumn{6}{l}{} \\\bottomrule[1pt]\addlinespace[-9pt]%
}%
\ctable[%
caption={Read first record data file: R-APDU},%
captionskip=-3pt,
label={tab:read+magstripe+r},%
pos={!ht},%
center,%
mincapwidth={\hsize},%
doinside={\footnotesize},%
width={\textwidth},%
]{p{9mm}p{26mm}llllll}{%
}{%
\toprule[1pt]
DATA & \ttfamily 70 6A 9F 6C 02 & \multicolumn{5}{l}{Tag: Non inter-industry nested data object template (\texttt{70})} \\%
     & \ttfamily 00 01 9F 62 06 & \multicolumn{5}{l}{Length: 106 (\texttt{6A})} \\%
     & \ttfamily 00 00 00 00 00 & \multicolumn{5}{l}{Value: (constructed)} \\%
     & \ttfamily 38 9F 63 06 00 & & & \multicolumn{4}{l}{Tag: Mag-Stripe application version number (\texttt{9F6C})} \\%
     & \ttfamily 00 00 00 03 C6 & & & \multicolumn{4}{l}{Length: 2 (\texttt{02})} \\%
     & \ttfamily 56 29 42 35 34 & & & \multicolumn{4}{l}{Value: Version 1 (\texttt{00 01})} \\%
     & \ttfamily 33 30 xx xx xx & & & \multicolumn{4}{l}{Tag: Track 1 bit map for CVC3 (\texttt{9F62})} \\%
     & \ttfamily xx 30 xx xx 37 & & & \multicolumn{4}{l}{Length: 6 (\texttt{06})} \\%
     & \ttfamily xx xx xx xx 5E & & & \multicolumn{4}{l}{Value: \texttt{00 00 00 00 00 38}} \\%
     & \ttfamily 20 2F 5E 31 37 & & & \multicolumn{4}{l}{Tag: Track 1 bit map for UN and ATC (\texttt{9F63})} \\%
     & \ttfamily 31 31 31 30 31 & & & \multicolumn{4}{l}{Length: 6 (\texttt{06})} \\%
     & \ttfamily 30 30 31 30 30 & & & \multicolumn{4}{l}{Value: \texttt{00 00 00 00 03 C6}} \\%
     & \ttfamily 30 30 30 30 30 & & & \multicolumn{4}{l}{Tag: Track 1 data (\texttt{56})} \\%
     & \ttfamily 30 30 30 9F 64 & & & \multicolumn{4}{l}{Length: 41 (\texttt{29})} \\%
     & \ttfamily 01 04 9F 65 02 & & & \multicolumn{4}{l}{Value: B5430xxxx0xx7xxxx\^{}\textvisiblespace/\^{}17111010010000000000} \\%
     & \ttfamily 00 38 9F 66 02 & & & \multicolumn{4}{l}{\ \ \ \ \ \ \ \ \ Format code: ``B'' (ISO/IEC 7813 Structure B)} \\%
     & \ttfamily 03 C6 9F 6B 13 & & & \multicolumn{4}{l}{\ \ \ \ \ \ \ \ \ PAN: ``5430 xxxx 0xx7 xxxx''} \\%
     & \ttfamily 54 30 xx xx 0x & & & \multicolumn{4}{l}{\ \ \ \ \ \ \ \ \ Field seperator: ``\^{}''} \\%
     & \ttfamily x7 xx xx D1 71 & & & \multicolumn{4}{l}{\ \ \ \ \ \ \ \ \ Cardholder: ``\textvisiblespace/''} \\%
     & \ttfamily 11 01 00 10 00 & & & \multicolumn{4}{l}{\ \ \ \ \ \ \ \ \ Field seperator: ``\^{}''} \\%
     & \ttfamily 00 00 00 0F 9F & & & \multicolumn{4}{l}{\ \ \ \ \ \ \ \ \ Expiry date: ``17''/``11''} \\%
     & \ttfamily 67 01 04       & & & \multicolumn{4}{l}{\ \ \ \ \ \ \ \ \ Service code: ``101''} \\%
     & \ttfamily                & & & \multicolumn{4}{l}{\ \ \ \ \ \ \ \ \ Discetionary data: ``0010000000000''} \\%
     & \ttfamily                & & & \multicolumn{4}{l}{Tag: Track 1 number of ATC digits (\texttt{9F64})} \\%
     & \ttfamily                & & & \multicolumn{4}{l}{Length: 1 (\texttt{01})} \\%
     & \ttfamily                & & & \multicolumn{4}{l}{Value: 4 (\texttt{04})} \\%
     & \ttfamily                & & & \multicolumn{4}{l}{Tag: Track 2 bit map for CVC3 (\texttt{9F65})} \\%
     & \ttfamily                & & & \multicolumn{4}{l}{Length: 2 (\texttt{02})} \\%
     & \ttfamily                & & & \multicolumn{4}{l}{Value: \texttt{00 38}} \\%
     & \ttfamily                & & & \multicolumn{4}{l}{Tag: Track 2 bit map for UN and ATC  (\texttt{9F66})} \\%
     & \ttfamily                & & & \multicolumn{4}{l}{Length: 2 (\texttt{02})} \\%
     & \ttfamily                & & & \multicolumn{4}{l}{Value: \texttt{03 C6}} \\%
     & \ttfamily                & & & \multicolumn{4}{l}{Tag: Track 2 data (\texttt{9F6B})} \\%
     & \ttfamily                & & & \multicolumn{4}{l}{Length: 19 (\texttt{13})} \\%
     & \ttfamily                & & & \multicolumn{4}{l}{Value: \texttt{5430xxxx0xx7xxxxD17111010010000000000F}} \\%
     & \ttfamily                & & & \multicolumn{4}{l}{\ \ \ \ \ \ \ \ \ PAN: ``5430 xxxx 0xx7 xxxx''} \\%
     & \ttfamily                & & & \multicolumn{4}{l}{\ \ \ \ \ \ \ \ \ Field seperator: ``D''} \\%
     & \ttfamily                & & & \multicolumn{4}{l}{\ \ \ \ \ \ \ \ \ Expiry date: ``17''/``11''} \\%
     & \ttfamily                & & & \multicolumn{4}{l}{\ \ \ \ \ \ \ \ \ Service code: ``101''} \\%
     & \ttfamily                & & & \multicolumn{4}{l}{\ \ \ \ \ \ \ \ \ Discetionary data: ``0010000000000''} \\%
     & \ttfamily                & & & \multicolumn{4}{l}{\ \ \ \ \ \ \ \ \ Padding: ``F''} \\%
     & \ttfamily                & & & \multicolumn{4}{l}{Tag: Track 2 number of ATC digits (\texttt{9F67})} \\%
     & \ttfamily                & & & \multicolumn{4}{l}{Length: 1 (\texttt{01})} \\%
     & \ttfamily                & & & \multicolumn{4}{l}{Value: 4 (\texttt{04})} \\\midrule[0.5pt]%
SW1  & \ttfamily 90 & \multicolumn{6}{l}{Success} \\%
SW2  & \ttfamily 00 & \multicolumn{6}{l}{} \\\bottomrule[1pt]\addlinespace[-9pt]%
}%

\ctable[%
caption={Compute Cryptographic Checksum: C-APDU},%
captionskip=-3pt,
label={tab:compute+cc+c},%
pos={!ht},%
center,%
mincapwidth={\hsize},%
doinside={\footnotesize},%
width={\textwidth},%
]{p{9mm}p{26mm}llllll}{%
}{%
\toprule[1pt]
CLA  & \ttfamily 80 & \multicolumn{6}{l}{Proprietary class} \\\midrule[0.5pt]%
INS  & \ttfamily 2A & \multicolumn{6}{l}{Compute Cryptographic Checksum} \\\midrule[0.5pt]%
P1   & \ttfamily 8E & \multicolumn{6}{l}{} \\\midrule[0.5pt]%
P2   & \ttfamily 80 & \multicolumn{6}{l}{} \\\midrule[0.5pt]%
Lc   & \ttfamily 04 & \multicolumn{6}{l}{} \\\midrule[0.5pt]%
DATA & \ttfamily 00 00 00 80 & \multicolumn{6}{l}{Unpredictable number (UN)} \\\midrule[0.5pt]%
Le   & \ttfamily 00 & \multicolumn{6}{l}{} \\\bottomrule[1pt]\addlinespace[-9pt]%
}%
\ctable[%
caption={Compute Cryptographic Checksum: R-APDU},%
captionskip=-3pt,
label={tab:compute+cc+r},%
pos={!ht},%
center,%
mincapwidth={\hsize},%
doinside={\footnotesize},%
width={\textwidth},%
]{p{9mm}p{26mm}llllll}{%
}{%
\toprule[1pt]
DATA & \ttfamily 77 0F 9F 61 02 & \multicolumn{5}{l}{Tag: Response message template (\texttt{77})} \\%
     & \ttfamily xx xx 9F 60 02 & \multicolumn{5}{l}{Length: 15 (\texttt{0F})} \\%
     & \ttfamily xx xx 9F 36 02 & \multicolumn{5}{l}{Value: (constructed)} \\%
     & \ttfamily 00 12          & & & \multicolumn{4}{l}{Tag: CVC3 Track 2 (\texttt{9F61})} \\%
     & \ttfamily                & & & \multicolumn{4}{l}{Length: 2 (\texttt{02})} \\%
     & \ttfamily                & & & \multicolumn{4}{l}{Value: \texttt{xx xx}} \\%
     & \ttfamily                & & & \multicolumn{4}{l}{Tag: CVC3 Track 1 (\texttt{9F60})} \\%
     & \ttfamily                & & & \multicolumn{4}{l}{Length: 2 (\texttt{02})} \\%
     & \ttfamily                & & & \multicolumn{4}{l}{Value: \texttt{xx xx}} \\%
     & \ttfamily                & & & \multicolumn{4}{l}{Tag: Application transaction counter (ATC) (\texttt{9F36})} \\%
     & \ttfamily                & & & \multicolumn{4}{l}{Length: 2 (\texttt{02})} \\%
     & \ttfamily                & & & \multicolumn{4}{l}{Value: 12 (\texttt{00 12})} \\\midrule[0.5pt]%
SW1  & \ttfamily 90 & \multicolumn{6}{l}{Success} \\%
SW2  & \ttfamily 00 & \multicolumn{6}{l}{} \\\bottomrule[1pt]\addlinespace[-9pt]%
}%

\clearpage
\section{Android Secure Element API}
\subsection{Class: NfcAdapterExtras}\label{appendix:nfcadapterextras}
\begin{lstlisting}[style=codeJAVA]
package com.android.nfc_extras;

public final class NfcAdapterExtras {
  /** Broadcast Action: RF field ON has been detected (unreliable/will be removed).  */
  public static final String ACTION_RF_FIELD_ON_DETECTED =
         "com.android.nfc_extras.action.RF_FIELD_ON_DETECTED";

  /** Broadcast Action: RF field OFF has been detected (unreliable/will be removed). */
  public static final String ACTION_RF_FIELD_OFF_DETECTED =
         "com.android.nfc_extras.action.RF_FIELD_OFF_DETECTED";


  /** Get the NfcAdapterExtras for the given NfcAdapter.                             */
  public static NfcAdapterExtras get(NfcAdapter adapter);


  /** Immutable data class that describes a card emulation route.                    */
  public final static class CardEmulationRoute {
    /** Card Emulation is turned off on this NfcAdapter.                             */
    public static final int ROUTE_OFF = 1;

    /** Card Emulation is routed to nfcEe when the screen is on,
      * otherwise it is turned off.                                                  */
    public static final int ROUTE_ON_WHEN_SCREEN_ON = 2;

    /** A route such as ROUTE_OFF or ROUTE_ON_WHEN_SCREEN_ON.                        */
    public final int route;

    /** The NFcExecutionEnvironment that Card Emulation is routed to.                */
    public final NfcExecutionEnvironment nfcEe;

    public CardEmulationRoute(int route, NfcExecutionEnvironment nfcEe);
  }


  /** Get the current routing state of the secure element.                           */
  public CardEmulationRoute getCardEmulationRoute();

  /** Set the routing state of the secure element.                                   */
  public void setCardEmulationRoute(CardEmulationRoute route);


  /** Get the NfcExecutionEnvironment for the embedded secure element.               */
  public NfcExecutionEnvironment getEmbeddedExecutionEnvironment();


  /** Authenticate the client application (if required by the implementation).
    * This method is not used on Nexus S/Galaxy Nexus.                               */
  public void authenticate(byte[] token);
}
\end{lstlisting}
\clearpage
\subsection{Class: NfcExecutionEnvironment}\label{appendix:nfcexecutionenvironment}
\begin{lstlisting}[style=codeJAVA]
package com.android.nfc_extras;

import java.io.IOException;

public class NfcExecutionEnvironment {
  /** Broadcast Action: An ISO-DEP AID was selected.                                 */
  public static final String ACTION_AID_SELECTED =
         "com.android.nfc_extras.action.AID_SELECTED";
  /** Mandatory byte array extra field in ACTION_AID_SELECTED.                       */
  public static final String EXTRA_AID = "com.android.nfc_extras.extra.AID";

  /** Broadcast action: A filtered APDU was received.                                */
  public static final String ACTION_APDU_RECEIVED =
         "com.android.nfc_extras.action.APDU_RECEIVED";
  /** Mandatory byte array extra field in ACTION_APDU_RECEIVED.                      */
  public static final String EXTRA_APDU_BYTES =
         "com.android.nfc_extras.extra.APDU_BYTES";

  /** Broadcast action: An EMV card removal event was detected.                      */
  public static final String ACTION_EMV_CARD_REMOVAL =
         "com.android.nfc_extras.action.EMV_CARD_REMOVAL";

  /** Broadcast action: An adapter implementing MIFARE Classic via card emulation
    * detected that a block has been accessed.                                       */
  public static final String ACTION_MIFARE_ACCESS_DETECTED =
         "com.android.nfc_extras.action.MIFARE_ACCESS_DETECTED";
  /** Optional integer extra field in ACTION_MIFARE_ACCESS_DETECTED that provides
    * the block number being accessed.                                               */
  public static final String EXTRA_MIFARE_BLOCK =
         "com.android.nfc_extras.extra.MIFARE_BLOCK";


  /** Open the NFC Execution Environment on its contact interface.                   */
  public void open() throws IOException;

  /** Close the NFC Execution Environment on its contact interface.                  */
  public void close() throws IOException;

  /** Send raw commands to the NFC-EE and receive the response.                      */
  public byte[] transceive(byte[] in) throws IOException;
}
\end{lstlisting}

\end{document}